\documentclass[aps,twocolumn,prl,showpacs,superscriptaddress]{revtex4-1}
\usepackage{graphicx}
\usepackage{amssymb} 
\usepackage{url}     
\usepackage{bm}      
\usepackage{color, soul}
\begin{document}

\newcommand{\ohm}{\ensuremath{\,\Omega}}
\newcommand{\kohm}{\ensuremath{\,\mbox{k}\Omega}}

\newcommand{\cb}{\ensuremath{C_{\mathrm{B}}}}
\newcommand{\ct}{\ensuremath{C_{\mathrm{T}}}}
\newcommand{\ntb}{\ensuremath{n_{\mathrm{T,B}}}}
\newcommand{\ctb}{\ensuremath{C_{\mathrm{T(B)}}}}
\newcommand{\cgg}{\ensuremath{C_{\mathrm{GG}}}}
\newcommand{\vsd}{\ensuremath{V_{\mathrm{SD}}}}
\newcommand{\vb}{\ensuremath{V_{\mathrm{B}}}}
\newcommand{\vt}{\ensuremath{V_{\mathrm{T}}}}
\newcommand{\vbg}{\ensuremath{V_{\mathrm{BG}}}}
\newcommand{\vtg}{\ensuremath{V_{\mathrm{TG}}}}
\newcommand{\vtgbg}{\ensuremath{V_{\mathrm{TG(BG)}}}}

\newcommand{\elale}{\ensuremath{\mathcal{E}_{\mathrm{LL}}}}

\newcommand{\nup}{\ensuremath{N_{\mathrm{U}}}}
\newcommand{\nlow}{\ensuremath{N_{\mathrm{L}}}}
\newcommand{\etot}{\ensuremath{E_{\mathrm{tot}}}}

\newcommand{\vtb}{\ensuremath{V_{\mathrm{T,B}}}}
\newcommand{\ntot}{\ensuremath{n_\mathrm{tot}}}
\newcommand{\nutot}{\ensuremath{\nu_\mathrm{tot}}}
\newcommand{\tnutot}{\ensuremath{\tilde{\nu}_\mathrm{tot}}}
\newcommand{\rxy}{\ensuremath{R_\mathrm{xy}}}
\newcommand{\rxx}{\ensuremath{R_\mathrm{xx}}}
\newcommand{\rhoxx}{\ensuremath{\rho_\mathrm{xx}}}
\newcommand{\rxxprime}{\ensuremath{R'_\mathrm{xx}}}
\newcommand{\egg}{\ensuremath{\epsilon_{\mathrm{GG}}}}
\newcommand{\dgg}{\ensuremath{d_{\mathrm{GG}}}}
\newcommand{\es}{\ensuremath{E_{\mathrm{S}}}}

\newcommand{\vbgo}{\ensuremath{V'_\mathrm{BG}}}
\newcommand{\vbgn}{\ensuremath{V^D_\mathrm{BG}}}
\newcommand{\vtgn}{\ensuremath{V^D_\mathrm{TG}}}
\newcommand{\nb}{\ensuremath{n_\mathrm{B}}}
\newcommand{\nt}{\ensuremath{n_\mathrm{T}}}
\newcommand{\ef}{\ensuremath{E_\mathrm{F}}}
\newcommand{\vbgbar}{\ensuremath{V_{\mathrm{BG}}- \vbgn}}
\newcommand{\vtgbar}{\ensuremath{V_\mathrm{TG}-\vtgn}}

\title{Quantum Hall Effect, Screening and Layer-Polarized Insulating States in Twisted Bilayer Graphene}
\author{Javier D Sanchez-Yamagishi}
\thanks{These authors contributed equally to this work}
\affiliation{Department of Physics, Massachusetts Institute of Technology, Cambridge, MA 02139 USA}
\author{Thiti  Taychatanapat}
\thanks{These authors contributed equally to this work}
\affiliation{Department of Physics, Harvard University, Cambridge, MA 02138 USA}
\author{Kenji Watanabe}
\affiliation{National Institute for Materials Science, Namiki 1-1, Tsukuba, Ibaraki 305-0044, Japan}
\author{Takashi Taniguchi}
\affiliation{National Institute for Materials Science, Namiki 1-1, Tsukuba, Ibaraki 305-0044, Japan}
\author{Amir Yacoby}
\affiliation{Department of Physics, Harvard University, Cambridge, MA 02138 USA}
\author{Pablo Jarillo-Herrero}
\affiliation{Department of Physics, Massachusetts Institute of Technology, Cambridge, MA 02139 USA}
\email{pjarillo@mit.edu}
\date{\today}
\begin{abstract}
We investigate electronic transport in dual-gated twisted bilayer graphene.  Despite the sub-nanometer proximity between the layers, we identify independent contributions to the magnetoresistance from the graphene Landau level spectrum of each layer.  We demonstrate that the filling factor of each layer can be independently controlled via the dual gates, which we use to induce Landau level crossings between the layers.  By analyzing the gate dependence of the Landau level crossings, we characterize the finite inter-layer screening and extract the capacitance between the atomically-spaced layers.  At zero filling factor, we observe magnetic and displacement field dependent insulating states, which indicate the presence of counter-propagating edge states with  inter-layer coupling.

\end{abstract}

\pacs{72.80.Vp, 73.22.Pr, 73.43.-f, 73.22.Gk,73.21.-b}
\maketitle

\begin{figure}[b]
\begin{center}
\includegraphics[width=3.375in]{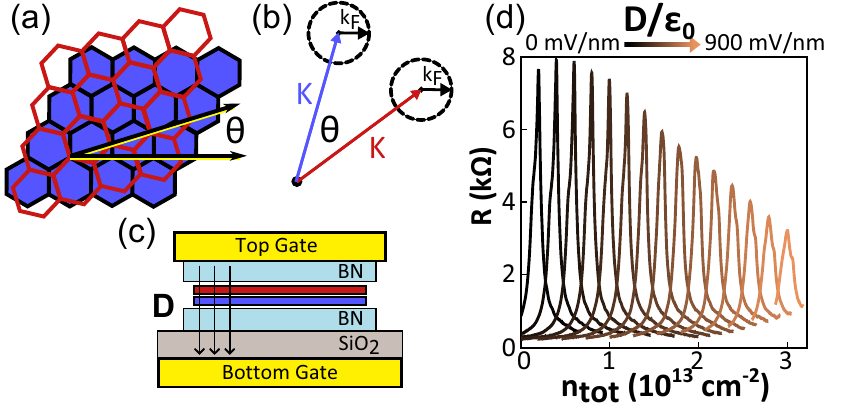}
\caption{ Twisted bilayer graphene device structure and zero magnetic field resistance measurements.  (a) Twisted bilayer graphene lattice with twist angle $\theta$. (b) Twist angle separates the Fermi surface of each layer in K-space.  (c) Schematic of a dual-gated twisted bilayer device with h-BN gate dielectric insulators.  Dual-gates allow for independent control of the carrier density and displacement field $D$. (d) Zero-magnetic field resistance $R$ at the charge neutrality point at different values of $D$.  The resistance at the charge neutrality point decreases with increasing $D$.  Peaks have been offset in density for clarity. }
\label{DeviceIntro}
\end{center}
\end{figure}

The bilayer 2-dimensional electron gas (2DEG) consists of two closely spaced 2DEGs, where inter-layer Coulomb interactions and tunneling effects can lead to new behaviors which are not present in the individual layers~\cite{Boebinger1990,Gramila1991,Eisenstein2004}.   In these bilayers, an insulating spacer is necessary to separate the 2DEG layers.  In the case of twisted bilayer graphene, the  layers can be stacked directly on top of each other, yet still retain a degree of independence.  This is possible because of the carbon honeycomb lattice of graphene, which results in weak coupling between the layers~\cite{Dresselhaus2002}, as well as a circular Fermi surface centered at nonzero K vectors~\cite{CastroNeto2009}.  The latter is key, because a relative twist angle between the graphene bilayer lattices can cause the Fermi surfaces of the two layers to not overlap at low densities (Fig.~\ref{DeviceIntro}a,b).  This preserves the linear Dirac dispersion in the twisted bilayer graphene~\cite{LopesdosSantos2007,Berger2006,Hass2008,Schmidt2008,Li2009,Luican2011,Schmidt2010}, but with twice the number of Dirac cones due to the two layers~\cite{LopesdosSantos2007,Schmidt2008,Schmidt2010}.  

\begin{figure*}
\includegraphics{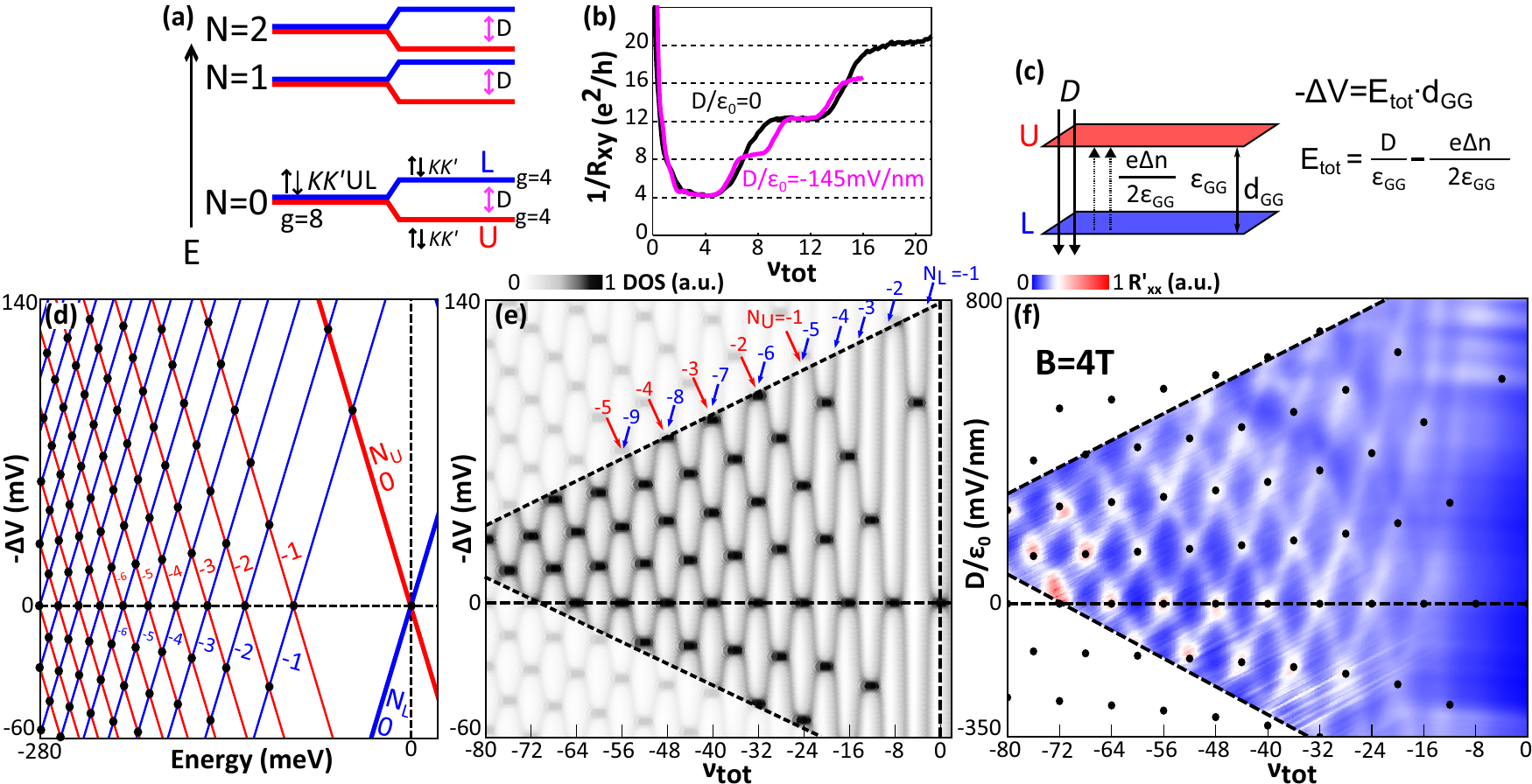} 
\caption{Quantum Hall effect, Landau level (LL) crossings, and screening in twisted bilayers.  (a) Schematic of twisted bilayer LL spectrum.  LLs are 8-fold degenerate (g=8) due to spin, valley \& layer degeneracy.  Displacement field $D$ breaks layer degeneracy (g=4).  (b) $1/\rxy$ as a function of total filling factor $\nutot$ at $B=9$T. At $D=0$, steps in $1/\rxy$ of $8e^2/h$  are observed (black line); at $D/\epsilon_0$=$-145$ mV/nm, new steps of $4e^2/h$ develop. (c) Diagram of inter-layer screening. The applied field $D$ is screened by charge imbalances $\Delta n$ and by the inter-layer dielectric constant $\egg$.  The total screened field $\etot$ induces an inter-layer potential difference $\Delta V$.  (d) LL energy spectra of upper and lower graphene layers (red and blue lines respectively) as a function of inter-layer potential difference $\Delta V$.  LL crossings are indicated by black dots.  $N_{U(L)}$ is the LL index of the upper~(lower) layer. (e) Simulated density of states for twisted bilayer as a function of $\nutot$ and $\Delta V$.  (f) Background-subtracted longitudinal resistance $\rxxprime$ as a function of $D$ and $\nutot$, measured at $B=4T$.  Peaks in $\rxxprime$ correspond to high density of states, where the Fermi level lies within a LL.  Black dots are theoretical fits to the LL crossings, from which the interlayer capacitance is extracted.} \label{QHEFigure}
\end{figure*}

Here, we present magnetoresistance measurements of dual-gated twisted bilayer graphene devices (twisted bilayers), which exhibit the quantum Hall effect (QHE) and magnetoresistance oscillations of two monolayer graphene (MLG) sheets conducting in parallel.  As we vary the gate voltages, we observe inter-layer Landau level crossings which allow us to quantify both the layer charge transfer as well as the finite screening effects between the layers. This incomplete screening of the applied field, due to graphene's small density of states and the close spacing between the layers, allows us to extract the inter-layer capacitance.  Lastly, at high magnetic fields we observe a pattern of insulating states centered at zero density which resemble those observed in AB-stacked bilayer graphene (AB-BLG)~\cite{Weitz2010,Kim2011}, but originate from layer-polarized edge modes.

Our twisted bilayer devices are fabricated using a PMMA-transfer technique to sequentially stack two separate MLG sheets such that they overlap on top of a hexagonal Boron Nitride (h-BN) flake~\cite{Dean2010,Taychatanapat2011}.  The bilayer region formed at the overlap is then contacted, and a topgate is fabricated with a h-BN flake as the dielectric insulator~\footnote{Additional details can be found in the Supplementary Information}.  The final devices are measured in a He3 cryostat, with the temperature at 300mK unless otherwise noted.

Using our devices' dual gates we can independently control the total carrier density $\ntot$ of the twisted bilayer, as well as the displacement field $D$ applied normal to the layers (Fig.~\ref{DeviceIntro}c).  The total carrier density of the twisted bilayer is $e\ntot=(\ct\vtg+\cb\vbg)$, where $\ctb$ is the capacitance per unit area to ground of the top~(bottom) gate, $\vtgbg$ is the potential difference between the top~(bottom) gate and the graphene layer closest to it~\footnotemark[1], and $e$ is the elementary charge.  The applied displacement field is $D=(\ct\vtg-\cb\vbg)/2$, which induces charge and voltage differences between the layers.

We first compare our twisted bilayer samples with AB-BLG by measuring the resistance of the charge neutrality point (CNP) as a function of $D$.  In AB-BLG, a displacement field breaks the bilayer's inversion symmetry, which opens a band gap at the CNP~\cite{McCann2006, Ohta2006, Oostinga2008a}.  This is not predicted to occur in twisted bilayers~\cite{LopesdosSantos2007}, and in our samples the CNP resistance decreases almost linearly as $D$ increases (Fig.~\ref{DeviceIntro}d).  This is a strong indication that our bilayers are not AB-stacked.  In this case, the effect of $D$ at the CNP is to dope the two layers with equal and opposite charge, reducing the resistance of each individual layer~\footnotemark[1].

At high magnetic field $B$, we measure a QHE which is distinctly different from that observed in MLG~\cite{Novoselov2005,Zhang2005} or AB-BLG~\cite{Novoselov2006}.  At $D=0$, we measure the Hall resistance $\rxy$ as a function of total filling factor $\nutot=\ntot h/eB$, where $h$ is Planck's constant  (Fig.~\ref{QHEFigure}b, black line).  We observe plateaus following the progression $1/\rxy=\nu(e^2/h)$, where $\nu=4,12,20$. These steps of $8 e^2/h$ between each plateau of $1/\rxy$  indicate the presence of 8-fold degenerate Landau levels (LLs).  This 8-fold degeneracy is due to the usual spin ($\uparrow$,$\downarrow$) and valley ($K$,$K'$) degeneracies found in MLG~\cite{Novoselov2005,Zhang2005}, with an additional 2-fold degeneracy  due to the layer degree of freedom (U,L for upper and lower layer respectively)(Fig.~\ref{QHEFigure}a)~\cite{DeGail2011}.
  
This layer degeneracy at $D=0$ was observed in three different samples, and can be seen up to high filling factors in the longitudinal resistance as well.  Fig.~\ref{QHEFigure}f shows longitudinal resistance measurements  $\rxxprime$, where a smooth background has been subtracted to improve the contrast of magnetoresistance peaks~\footnotemark[1].  When $D=0$, peaks in $\rxxprime$ are separated by $\Delta\nutot=8$, due to the 8-fold degeneracy, with this trend verified as far as $\nutot=-72$.  

A property of the twisted bilayers is that the layer degeneracy can be easily broken by applying a displacement field normal to the graphene layers, resulting in 4-fold degenerate LLs with steps of $4 e^2/h$ in $1/\rxy$ (Fig.~\ref{QHEFigure}b, purple line).    These LL splittings are also seen clearly in $\rxxprime$, where each peak in $\rxxprime$ at $D=0$ splits in two for $|D|>0$ (Fig.~\ref{QHEFigure}f).  As $D$ is increased further, these peaks cross with their neighbors, corresponding to the crossing of LLs between the layers. 

To model the pattern of possible LL crossings, we consider independent MLG LL energy spectra in each layer with a potential difference $\Delta V$ between the upper and lower layer induced by $D$ (Fig.~\ref{QHEFigure}d).  The upper and lower layer LLs (red and blue lines respectively), are degenerate at $\Delta V=0$, and split in energy as $|\Delta V|$ increases, resulting in energy crossings when $-e\Delta V$ is equal to the energy spacing between two MLG LLs.  This condition is satisfied when $-e\Delta V=\elale(\nup)-\elale(\nlow)$, where $\elale(N)=\text{sgn}(N)v_F \sqrt{2e\hbar B|N|}$, $\nup$ and $\nlow$ are the LL indices for the upper and lower layer respectively, and $v_F$ is the MLG Fermi velocity~\cite{CastroNeto2009}.  This energy plot is converted to filling factor by modeling each LL by a Lorentzian density of states with disorder broadening (Fig.~\ref{QHEFigure}e).   The resulting plot of two intersecting LL spectra qualitatively reproduces all the peaks in $\rxxprime$ presented in Fig.~\ref{QHEFigure}f.  

\begin{figure}[b]
\includegraphics{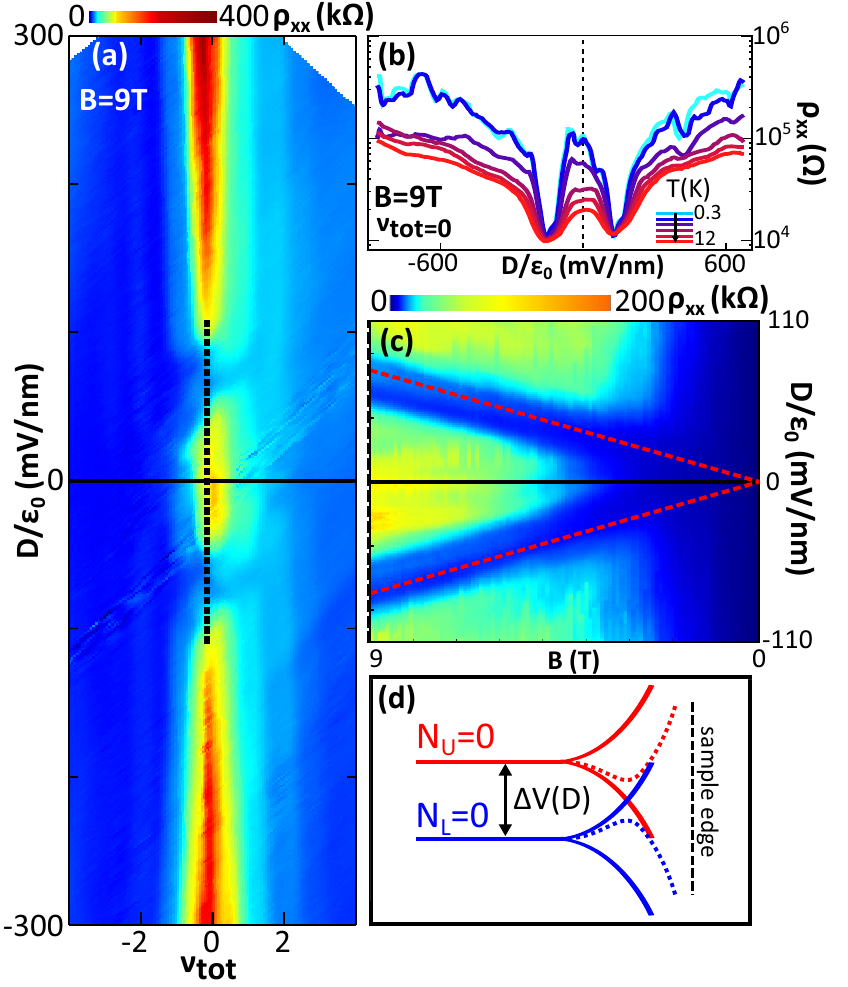} 
\caption{Insulating states in twisted bilayer at $\nutot=0$. (a) Longitudinal resistivity $\rhoxx$ as a function of $D$ and $\nutot$ at $B$=9T. At $\nutot=0$, two insulating regimes are observed, one at $D=0$ and another at high $D$, with a low $\rhoxx$ region separating them. (b)  Temperature dependence at $B$=9T and $\nutot=0$ of $\rhoxx$  vs $D$ shows non-metallic behavior.  Temperature increases going from the top light blue curve to bottom red curve as 0.3, 1, 4, 8, 10, and 12K, respectively.  (c) Magnetic field dependence of $\nutot=0$ insulating states.  Resistivity double minima approach each other with slope 7.5 mV/nmT (dashed red lines).  Both insulating states disappear at low $B$.   (d) Schematic of $\nutot=0$ edge states at nonzero $\Delta V$ when $D$ is applied.  The zeroth LLs are split apart, resulting in counter-propagating edge states in the absence of interactions (intersecting solid lines).  The insulating state at high $D$ indicates inter-layer coupling between these edge states, which may open a gap at the edge (dashed lines).} \label{NuZero}
\end{figure}
  The relationship between the applied $D$ and the induced $\Delta V$ at a crossing is determined by the inter-layer screening properties of the twisted bilayer, as $D$ will be screened both by free charges and the inter-layer dielectric environment  (Fig.~\ref{QHEFigure}c). The total screened electric field $\etot$ between the two graphene sheets with spacing $\dgg$ results in the potential difference $-\Delta V= {\etot} \cdot \dgg$. The relation then is:

\begin{equation}
-\Delta V = \left ( D-{e \Delta n \over 2}   \right ) {\dgg \over {\egg}}= \left ( D-\epsilon_0 \es\right ) {{1}\over{\cgg}}, \label{ScreeningEquation}
\end{equation}
where $\es$ is the screening field due to the layer density imbalance $\Delta n$, $\egg$  is the inter-layer dielectric constant, and $\cgg=\egg/\dgg$ is the inter-layer capacitance per unit area.

For a high density of states material, $\Delta V$ would be effectively zero and the charge imbalance $\Delta n$ would completely screen $D$, independent of the inter-layer capacitance $\cgg$.  Graphene though, has a small density of states, and a correspondingly small quantum capacitance which is comparable to the inter-layer capacitance of the closely-spaced graphene sheets.  This leads to an incomplete charge screening of $D$ and a dependence of $\Delta n$ on $\cgg$ which we can measure. When the Fermi energy lies at a LL crossing with LL indices $\nup$ and $\nlow$, we can determine both $\Delta V$ and $\Delta n$ ($\Delta n=(\nup-\nlow)4eB/h$) and use equation (\ref{ScreeningEquation}) to compute the $D$ at which a crossing should be observed.  We repeat this process for each crossing, and fit it to our data to extract $\cgg$.  The computed crossings are overlaid as black circles on Fig.~\ref{QHEFigure}f, resulting in good agreement when $\cgg=6.8~\mu\text{F/cm}^2$ (estimated error $\pm 1.0~\mu\text{F/cm}^2$).  A similar analysis was repeated on two other samples with LL crossings, both resulting in an extracted capacitance of $\cgg=7.5\pm1.0~\mu \text{F/cm}^2$.  For comparison, the capacitance of two parallel plates separated by $0.34$~nm of vacuum would be $2.6~\mu  \text{F/cm}^2$, which is less than half of our extracted capacitances (atomic force microscopy measurements indicate an inter-layer step height that varies from 3.4 to 4.1~$\text{\AA}$ across our samples).  Given that the inter-layer distances are only somewhat larger than the spatial extent of the graphene $p_z$ orbitals~\cite{Huang2006}, it seems likely that both the finite thickness of the graphene layer, and its polarizability~\cite{PhysRevB.75.205418}, could increase the inter-layer capacitance.  Another possible effect is Fermi velocity reduction, which has been demonstrated to occur in twisted bilayers at small twist angles ~\cite{LopesdosSantos2007,Shallcross2010,Luican2011,Schmidt2010}.  In this case, since we assume $v_F$ to be the same as in isolated MLG, the inter-layer capacitance could be even larger than we estimate and our extracted capacitance $\cgg$ sets a lower-bound on this quantity.

We now turn to the $\nutot=0$ region, where we see evidence of coupling between edge states in the two layers.  At high B-field and $\nutot=0$, the longitudinal resistivity $\rhoxx$  has two insulating regions: one at $D=0$ and one at high $D$ (Fig.~\ref{NuZero}a).  Both insulating states have high resistivities ($>100~\text{k}\Omega$) with a non-metallic temperature dependence (Fig.~\ref{NuZero}b). A similar pattern of insulating states has been observed in AB-BLG~\cite{Weitz2010,Kim2011}, but the mechanism for such states must be different in the twisted bilayers.  In AB-BLG, high $D$ opens a band gap independent of $B$.  Such an effect does not occur in the twisted bilayers (Fig.~\ref{DeviceIntro}d)~\cite{LopesdosSantos2007}, and as seen in Fig.~\ref{NuZero}c, the high $D$ insulating state disappears at low B field.

This high $D$ insulating state can be explained by the coupling of counter-propagating edge states, which can co-exist on the same edge of the twisted-bilayer sample when $\nutot=0$.  These crossings occur when $|D|>0$, because the zeroth LL in graphene is made up of opposite chirality states with energy that diverges in opposite directions at the edge of the sample (Fig.~\ref{NuZero}d).  When the zeroth LL of the twisted bilayer is split in energy by $D$, the electron-like edge states in one layer (blue line) will cross the hole-like edge states in the other (red line), resulting in counter-propagating, layer-polarized edge modes.  A similar scenario has been previously considered for spin-splitting in the zeroth LL in graphene, leading to spin currents~\cite{Abanin2006}.  In the case of twisted bilayer though, there should be a displacement-induced layer splitting of the zeroth LL, with associated ``layer" current.  Because the states counter-propagate along the same edge though, a backscattering channel is available by tunneling into the other layer.  Such a process could lead to 1d localization~\cite{Lee1985}, or an insulating gap due to an avoided crossing of the edges states (Fig.~\ref{NuZero}d, dotted lines), both of which could explain the insulating behavior we observe.

The high $D$ insulating state at $ \nutot=0$ is separated by a low resistivity region from another insulating state at $D=0$.  The development of an insulating state at zero filling factor has been observed in MLG~\cite{Zhang2006,*Checkelsky2009,*Giesbers2009} and BLG~\cite{Zhao2010, Weitz2010}, and is attributed to electron-electron interaction effects which break the degeneracy of the zeroth Landau level and open a gap at zero density.  Given the presence of the low resistivity transition region between the two insulating states, it is unlikely then that both the high $D$ and $D=0$ regions are layer-polarized states, since that would imply a continuous transition from one state to the other.  The $D=0$ state then could simply be both MLG sheets within some broken-symmetry state that does not involve the layer degree of freedom~\cite{Kharitonov2011}. As $D$ increases, the layer-polarized state eventually becomes more energetically favorable, leading to the transition to the high $D$ insulating state.

These $\nutot=0$ states indicate that layer interactions in the twisted bilayer graphene can lead to new behaviors which cannot be explained by completely independent monolayer graphene sheets conducting in parallel.  In principle, this inter-layer coupling is tunable by varying the distance between the graphene layers, altering the twist angle, as well as by threading magnetic flux parallel to the layers.

Research supported by the US Department of Energy, Office of Basic Energy Sciences, Division of Materials Sciences and Engineering under Award DE-SC0001819. Sample fabrication was performed partly at the NSF funded MIT Center for Materials Science and Engineering and Harvard Center for Nanoscale Science.  We thank M.Y. Kharitonov, M.M. Fogler, F. Guinea, R. Nandkishore and A.F. Young for discussion.  


\begin{thebibliography}{34}%
\makeatletter
\providecommand \@ifxundefined [1]{%
 \@ifx{#1\undefined}
}%
\providecommand \@ifnum [1]{%
 \ifnum #1\expandafter \@firstoftwo
 \else \expandafter \@secondoftwo
 \fi
}%
\providecommand \@ifx [1]{%
 \ifx #1\expandafter \@firstoftwo
 \else \expandafter \@secondoftwo
 \fi
}%
\providecommand \natexlab [1]{#1}%
\providecommand \enquote  [1]{``#1''}%
\providecommand \bibnamefont  [1]{#1}%
\providecommand \bibfnamefont [1]{#1}%
\providecommand \citenamefont [1]{#1}%
\providecommand \href@noop [0]{\@secondoftwo}%
\providecommand \href [0]{\begingroup \@sanitize@url \@href}%
\providecommand \@href[1]{\@@startlink{#1}\@@href}%
\providecommand \@@href[1]{\endgroup#1\@@endlink}%
\providecommand \@sanitize@url [0]{\catcode `\\12\catcode `\$12\catcode
  `\&12\catcode `\#12\catcode `\^12\catcode `\_12\catcode `\%12\relax}%
\providecommand \@@startlink[1]{}%
\providecommand \@@endlink[0]{}%
\providecommand \url  [0]{\begingroup\@sanitize@url \@url }%
\providecommand \@url [1]{\endgroup\@href {#1}{\urlprefix }}%
\providecommand \urlprefix  [0]{URL }%
\providecommand \Eprint [0]{\href }%
\providecommand \doibase [0]{http://dx.doi.org/}%
\providecommand \selectlanguage [0]{\@gobble}%
\providecommand \bibinfo  [0]{\@secondoftwo}%
\providecommand \bibfield  [0]{\@secondoftwo}%
\providecommand \translation [1]{[#1]}%
\providecommand \BibitemOpen [0]{}%
\providecommand \bibitemStop [0]{}%
\providecommand \bibitemNoStop [0]{.\EOS\space}%
\providecommand \EOS [0]{\spacefactor3000\relax}%
\providecommand \BibitemShut  [1]{\csname bibitem#1\endcsname}%
\let\auto@bib@innerbib\@empty
\bibitem [{\citenamefont {Boebinger}\ \emph {et~al.}(1990)\citenamefont
  {Boebinger}, \citenamefont {Jiang}, \citenamefont {Pfeiffer},\ and\
  \citenamefont {West}}]{Boebinger1990}%
  \BibitemOpen
  \bibfield  {author} {\bibinfo {author} {\bibfnamefont {G.~S.}\ \bibnamefont
  {Boebinger}}, \bibinfo {author} {\bibfnamefont {H.~W.}\ \bibnamefont
  {Jiang}}, \bibinfo {author} {\bibfnamefont {L.~N.}\ \bibnamefont {Pfeiffer}},
  \ and\ \bibinfo {author} {\bibfnamefont {K.~W.}\ \bibnamefont {West}},\
  }\href {\doibase 10.1103/PhysRevLett.64.1793} {\bibfield  {journal} {\bibinfo
   {journal} {Phys. Rev. Lett.}\ }\textbf {\bibinfo {volume} {64}},\ \bibinfo
  {pages} {1793} (\bibinfo {year} {1990})}\BibitemShut {NoStop}%
\bibitem [{\citenamefont {Gramila}\ \emph {et~al.}(1991)\citenamefont {Gramila}
  \emph {et~al.}}]{Gramila1991}%
  \BibitemOpen
  \bibfield  {author} {\bibinfo {author} {\bibfnamefont {T.~J.}\ \bibnamefont
  {Gramila}} \emph {et~al.},\ }\href {\doibase 10.1103/PhysRevLett.66.1216}
  {\bibfield  {journal} {\bibinfo  {journal} {Phys. Rev. Lett.}\ }\textbf
  {\bibinfo {volume} {66}},\ \bibinfo {pages} {1216} (\bibinfo {year}
  {1991})}\BibitemShut {NoStop}%
\bibitem [{\citenamefont {Eisenstein}\ and\ \citenamefont
  {Macdonald}(2004)}]{Eisenstein2004}%
  \BibitemOpen
  \bibfield  {author} {\bibinfo {author} {\bibfnamefont {J.~P.}\ \bibnamefont
  {Eisenstein}}\ and\ \bibinfo {author} {\bibfnamefont {A.~H.}\ \bibnamefont
  {Macdonald}},\ }\href {\doibase 10.1038/nature03081} {\bibfield  {journal}
  {\bibinfo  {journal} {Nature}\ }\textbf {\bibinfo {volume} {432}},\ \bibinfo
  {pages} {691} (\bibinfo {year} {2004})}\BibitemShut {NoStop}%
\bibitem [{\citenamefont {Dresselhaus}\ and\ \citenamefont
  {Dresselhaus}(2002)}]{Dresselhaus2002}%
  \BibitemOpen
  \bibfield  {author} {\bibinfo {author} {\bibfnamefont {M.~S.}\ \bibnamefont
  {Dresselhaus}}\ and\ \bibinfo {author} {\bibfnamefont {G.}~\bibnamefont
  {Dresselhaus}},\ }\href {\doibase 10.1080/00018730110113644} {\bibfield
  {journal} {\bibinfo  {journal} {Adv. Phys.}\ }\textbf {\bibinfo {volume}
  {51}},\ \bibinfo {pages} {1} (\bibinfo {year} {2002})}\BibitemShut {NoStop}%
\bibitem [{\citenamefont {Castro~Neto}\ \emph {et~al.}(2009)\citenamefont
  {Castro~Neto} \emph {et~al.}}]{CastroNeto2009}%
  \BibitemOpen
  \bibfield  {author} {\bibinfo {author} {\bibfnamefont {A.~H.}\ \bibnamefont
  {Castro~Neto}} \emph {et~al.},\ }\href {\doibase 10.1103/RevModPhys.81.109}
  {\bibfield  {journal} {\bibinfo  {journal} {Rev. Mod. Phys.}\ }\textbf
  {\bibinfo {volume} {81}},\ \bibinfo {pages} {109} (\bibinfo {year}
  {2009})}\BibitemShut {NoStop}%
\bibitem [{\citenamefont {{Lopes dos Santos}}\ \emph
  {et~al.}(2007)\citenamefont {{Lopes dos Santos}}, \citenamefont {Peres},\
  and\ \citenamefont {{Castro Neto}}}]{LopesdosSantos2007}%
  \BibitemOpen
  \bibfield  {author} {\bibinfo {author} {\bibfnamefont {J.~M.~B.}\
  \bibnamefont {{Lopes dos Santos}}}, \bibinfo {author} {\bibfnamefont
  {N.~M.~R.}\ \bibnamefont {Peres}}, \ and\ \bibinfo {author} {\bibfnamefont
  {A.~H.}\ \bibnamefont {{Castro Neto}}},\ }\href {\doibase
  10.1103/PhysRevLett.99.256802} {\bibfield  {journal} {\bibinfo  {journal}
  {Phys. Rev. Lett.}\ }\textbf {\bibinfo {volume} {99}},\ \bibinfo {pages}
  {256802} (\bibinfo {year} {2007})}\BibitemShut {NoStop}%
\bibitem [{\citenamefont {{C. Berger \emph{et~al.}}}(2006)}]{Berger2006}%
  \BibitemOpen
  \bibfield  {author} {\bibinfo {author} {\bibnamefont {{C. Berger
  \emph{et~al.}}}},\ }\href {\doibase 10.1126/science.1125925} {\bibfield
  {journal} {\bibinfo  {journal} {Science}\ }\textbf {\bibinfo {volume}
  {312}},\ \bibinfo {pages} {1191} (\bibinfo {year} {2006})}\BibitemShut
  {NoStop}%
\bibitem [{\citenamefont {Hass}\ \emph {et~al.}(2008)\citenamefont {Hass} \emph
  {et~al.}}]{Hass2008}%
  \BibitemOpen
  \bibfield  {author} {\bibinfo {author} {\bibfnamefont {J.}~\bibnamefont
  {Hass}} \emph {et~al.},\ }\href {\doibase 10.1103/PhysRevLett.100.125504}
  {\bibfield  {journal} {\bibinfo  {journal} {Phys. Rev. Lett.}\ }\textbf
  {\bibinfo {volume} {100}},\ \bibinfo {pages} {125504} (\bibinfo {year}
  {2008})}\BibitemShut {NoStop}%
\bibitem [{\citenamefont {Schmidt}\ \emph {et~al.}(2008)\citenamefont {Schmidt}
  \emph {et~al.}}]{Schmidt2008}%
  \BibitemOpen
  \bibfield  {author} {\bibinfo {author} {\bibfnamefont {H.}~\bibnamefont
  {Schmidt}} \emph {et~al.},\ }\href
  {http://link.aip.org/link/doi/10.1063/1.3012369/html} {\bibfield  {journal}
  {\bibinfo  {journal} {Appl. Phys. Lett.}\ }\textbf {\bibinfo {volume} {93}},\
  \bibinfo {pages} {172108} (\bibinfo {year} {2008})}\BibitemShut {NoStop}%
\bibitem [{\citenamefont {Li}\ \emph {et~al.}(2009)\citenamefont {Li} \emph
  {et~al.}}]{Li2009}%
  \BibitemOpen
  \bibfield  {author} {\bibinfo {author} {\bibfnamefont {G.}~\bibnamefont {Li}}
  \emph {et~al.},\ }\href {\doibase 10.1038/nphys1463} {\bibfield  {journal}
  {\bibinfo  {journal} {Nat. Phys.}\ }\textbf {\bibinfo {volume} {6}},\
  \bibinfo {pages} {109} (\bibinfo {year} {2009})}\BibitemShut {NoStop}%
\bibitem [{\citenamefont {Luican}\ \emph {et~al.}(2011)\citenamefont {Luican}
  \emph {et~al.}}]{Luican2011}%
  \BibitemOpen
  \bibfield  {author} {\bibinfo {author} {\bibfnamefont {A.}~\bibnamefont
  {Luican}} \emph {et~al.},\ }\href
  {http://link.aps.org/doi/10.1103/PhysRevLett.106.126802} {\bibfield
  {journal} {\bibinfo  {journal} {Phys. Rev. Lett.}\ }\textbf {\bibinfo
  {volume} {106}},\ \bibinfo {pages} {126802} (\bibinfo {year}
  {2011})}\BibitemShut {NoStop}%
\bibitem [{\citenamefont {Schmidt}\ \emph {et~al.}(2010)\citenamefont
  {Schmidt}, \citenamefont {L\"udtke}, \citenamefont {Barthold},\ and\
  \citenamefont {Haug}}]{Schmidt2010}%
  \BibitemOpen
  \bibfield  {author} {\bibinfo {author} {\bibfnamefont {H.}~\bibnamefont
  {Schmidt}}, \bibinfo {author} {\bibfnamefont {T.}~\bibnamefont {L\"udtke}},
  \bibinfo {author} {\bibfnamefont {P.}~\bibnamefont {Barthold}}, \ and\
  \bibinfo {author} {\bibfnamefont {R.~J.}\ \bibnamefont {Haug}},\ }\href
  {http://www.sciencedirect.com/science/article/pii/S1386947709005803}
  {\bibfield  {journal} {\bibinfo  {journal} {Physica E}\ }\textbf {\bibinfo
  {volume} {42}},\ \bibinfo {pages} {699} (\bibinfo {year} {2010})}\BibitemShut
  {NoStop}%
\bibitem [{\citenamefont {Weitz}\ \emph {et~al.}(2010)\citenamefont {Weitz}
  \emph {et~al.}}]{Weitz2010}%
  \BibitemOpen
  \bibfield  {author} {\bibinfo {author} {\bibfnamefont {R.~T.}\ \bibnamefont
  {Weitz}} \emph {et~al.},\ }\href {\doibase 10.1126/science.1194988}
  {\bibfield  {journal} {\bibinfo  {journal} {Science}\ }\textbf {\bibinfo
  {volume} {330}},\ \bibinfo {pages} {812} (\bibinfo {year}
  {2010})}\BibitemShut {NoStop}%
\bibitem [{\citenamefont {Kim}\ \emph {et~al.}(2011)\citenamefont {Kim},
  \citenamefont {Lee},\ and\ \citenamefont {Tutuc}}]{Kim2011}%
  \BibitemOpen
  \bibfield  {author} {\bibinfo {author} {\bibfnamefont {S.}~\bibnamefont
  {Kim}}, \bibinfo {author} {\bibfnamefont {K.}~\bibnamefont {Lee}}, \ and\
  \bibinfo {author} {\bibfnamefont {E.}~\bibnamefont {Tutuc}},\ }\href
  {http://link.aps.org/doi/10.1103/PhysRevLett.107.016803} {\bibfield
  {journal} {\bibinfo  {journal} {Phys. Rev. Lett.}\ }\textbf {\bibinfo
  {volume} {107}},\ \bibinfo {pages} {16803} (\bibinfo {year}
  {2011})}\BibitemShut {NoStop}%
\bibitem [{\citenamefont {Dean}\ \emph {et~al.}(2010)\citenamefont {Dean} \emph
  {et~al.}}]{Dean2010}%
  \BibitemOpen
  \bibfield  {author} {\bibinfo {author} {\bibfnamefont {C.~R.}\ \bibnamefont
  {Dean}} \emph {et~al.},\ }\href {\doibase 10.1038/nnano.2010.172} {\bibfield
  {journal} {\bibinfo  {journal} {Nat. Nanotechnol.}\ }\textbf {\bibinfo
  {volume} {5}},\ \bibinfo {pages} {722} (\bibinfo {year} {2010})}\BibitemShut
  {NoStop}%
\bibitem [{\citenamefont {Taychatanapat}\ \emph {et~al.}(2011)\citenamefont
  {Taychatanapat}, \citenamefont {Watanabe}, \citenamefont {Taniguchi},\ and\
  \citenamefont {Jarillo-Herrero}}]{Taychatanapat2011}%
  \BibitemOpen
  \bibfield  {author} {\bibinfo {author} {\bibfnamefont {T.}~\bibnamefont
  {Taychatanapat}}, \bibinfo {author} {\bibfnamefont {K.}~\bibnamefont
  {Watanabe}}, \bibinfo {author} {\bibfnamefont {T.}~\bibnamefont {Taniguchi}},
  \ and\ \bibinfo {author} {\bibfnamefont {P.}~\bibnamefont
  {Jarillo-Herrero}},\ }\href {\doibase 10.1038/nphys2008} {\bibfield
  {journal} {\bibinfo  {journal} {Nat. Phys.}\ }\textbf {\bibinfo {volume}
  {7}},\ \bibinfo {pages} {621} (\bibinfo {year} {2011})}\BibitemShut {NoStop}%
\bibitem [{Note1()}]{Note1}%
  \BibitemOpen
  \bibinfo {note} {Additional details can be found in the Supplementary
  Information}\BibitemShut {NoStop}%
\bibitem [{\citenamefont {McCann}(2006)}]{McCann2006}%
  \BibitemOpen
  \bibfield  {author} {\bibinfo {author} {\bibfnamefont {E.}~\bibnamefont
  {McCann}},\ }\href {\doibase 10.1103/PhysRevB.74.161403} {\bibfield
  {journal} {\bibinfo  {journal} {Phys. Rev. B}\ }\textbf {\bibinfo {volume}
  {74}},\ \bibinfo {pages} {161403} (\bibinfo {year} {2006})}\BibitemShut
  {NoStop}%
\bibitem [{\citenamefont {Ohta}\ \emph {et~al.}(2006)\citenamefont {Ohta} \emph
  {et~al.}}]{Ohta2006}%
  \BibitemOpen
  \bibfield  {author} {\bibinfo {author} {\bibfnamefont {T.}~\bibnamefont
  {Ohta}} \emph {et~al.},\ }\href {\doibase 10.1126/science.1130681} {\bibfield
   {journal} {\bibinfo  {journal} {Science}\ }\textbf {\bibinfo {volume}
  {313}},\ \bibinfo {pages} {951} (\bibinfo {year} {2006})}\BibitemShut
  {NoStop}%
\bibitem [{\citenamefont {Oostinga}\ \emph {et~al.}(2007)\citenamefont
  {Oostinga} \emph {et~al.}}]{Oostinga2008a}%
  \BibitemOpen
  \bibfield  {author} {\bibinfo {author} {\bibfnamefont {J.~B.}\ \bibnamefont
  {Oostinga}} \emph {et~al.},\ }\href {\doibase 10.1038/nmat2082} {\bibfield
  {journal} {\bibinfo  {journal} {Nat. Mater.}\ }\textbf {\bibinfo {volume}
  {7}},\ \bibinfo {pages} {151} (\bibinfo {year} {2007})}\BibitemShut {NoStop}%
\bibitem [{\citenamefont {Novoselov}\ \emph {et~al.}(2005)\citenamefont
  {Novoselov} \emph {et~al.}}]{Novoselov2005}%
  \BibitemOpen
  \bibfield  {author} {\bibinfo {author} {\bibfnamefont {K.~S.}\ \bibnamefont
  {Novoselov}} \emph {et~al.},\ }\href {\doibase 10.1038/nature04233}
  {\bibfield  {journal} {\bibinfo  {journal} {Nature}\ }\textbf {\bibinfo
  {volume} {438}},\ \bibinfo {pages} {197} (\bibinfo {year}
  {2005})}\BibitemShut {NoStop}%
\bibitem [{\citenamefont {Zhang}\ \emph {et~al.}(2005)\citenamefont {Zhang},
  \citenamefont {Tan}, \citenamefont {Stormer},\ and\ \citenamefont
  {Kim}}]{Zhang2005}%
  \BibitemOpen
  \bibfield  {author} {\bibinfo {author} {\bibfnamefont {Y.}~\bibnamefont
  {Zhang}}, \bibinfo {author} {\bibfnamefont {Y.-W.}\ \bibnamefont {Tan}},
  \bibinfo {author} {\bibfnamefont {H.~L.}\ \bibnamefont {Stormer}}, \ and\
  \bibinfo {author} {\bibfnamefont {P.}~\bibnamefont {Kim}},\ }\href {\doibase
  10.1038/nature04235} {\bibfield  {journal} {\bibinfo  {journal} {Nature}\
  }\textbf {\bibinfo {volume} {438}},\ \bibinfo {pages} {201} (\bibinfo {year}
  {2005})}\BibitemShut {NoStop}%
\bibitem [{\citenamefont {Novoselov}\ \emph {et~al.}(2006)\citenamefont
  {Novoselov} \emph {et~al.}}]{Novoselov2006}%
  \BibitemOpen
  \bibfield  {author} {\bibinfo {author} {\bibfnamefont {K.~S.}\ \bibnamefont
  {Novoselov}} \emph {et~al.},\ }\href {\doibase 10.1038/nphys245} {\bibfield
  {journal} {\bibinfo  {journal} {Nat. Phys.}\ }\textbf {\bibinfo {volume}
  {2}},\ \bibinfo {pages} {177} (\bibinfo {year} {2006})}\BibitemShut {NoStop}%
\bibitem [{\citenamefont {de~Gail}\ \emph {et~al.}(2011)\citenamefont {de~Gail}
  \emph {et~al.}}]{DeGail2011}%
  \BibitemOpen
  \bibfield  {author} {\bibinfo {author} {\bibfnamefont {R.}~\bibnamefont
  {de~Gail}} \emph {et~al.},\ }\href {\doibase 10.1103/PhysRevB.84.045436}
  {\bibfield  {journal} {\bibinfo  {journal} {Phys. Rev. B}\ }\textbf {\bibinfo
  {volume} {84}},\ \bibinfo {pages} {045436} (\bibinfo {year}
  {2011})}\BibitemShut {NoStop}%
\bibitem [{\citenamefont {Huang}\ \emph {et~al.}(2006)\citenamefont {Huang},
  \citenamefont {Wu},\ and\ \citenamefont {Hwang}}]{Huang2006}%
  \BibitemOpen
  \bibfield  {author} {\bibinfo {author} {\bibfnamefont {Y.}~\bibnamefont
  {Huang}}, \bibinfo {author} {\bibfnamefont {J.}~\bibnamefont {Wu}}, \ and\
  \bibinfo {author} {\bibfnamefont {K.~C.}\ \bibnamefont {Hwang}},\ }\href
  {\doibase 10.1103/PhysRevB.74.245413} {\bibfield  {journal} {\bibinfo
  {journal} {Phys. Rev. B}\ }\textbf {\bibinfo {volume} {74}},\ \bibinfo
  {pages} {245413} (\bibinfo {year} {2006})}\BibitemShut {NoStop}%
\bibitem [{\citenamefont {Hwang}\ and\ \citenamefont {{Das
  Sarma}}(2007)}]{PhysRevB.75.205418}%
  \BibitemOpen
  \bibfield  {author} {\bibinfo {author} {\bibfnamefont {E.~H.}\ \bibnamefont
  {Hwang}}\ and\ \bibinfo {author} {\bibfnamefont {S.}~\bibnamefont {{Das
  Sarma}}},\ }\href {\doibase 10.1103/PhysRevB.75.205418} {\bibfield  {journal}
  {\bibinfo  {journal} {Phys. Rev. B}\ }\textbf {\bibinfo {volume} {75}},\
  \bibinfo {pages} {205418} (\bibinfo {year} {2007})}\BibitemShut {NoStop}%
\bibitem [{\citenamefont {Shallcross}\ \emph {et~al.}(2010)\citenamefont
  {Shallcross}, \citenamefont {Sharma}, \citenamefont {Kandelaki},\ and\
  \citenamefont {Pankratov}}]{Shallcross2010}%
  \BibitemOpen
  \bibfield  {author} {\bibinfo {author} {\bibfnamefont {S.}~\bibnamefont
  {Shallcross}}, \bibinfo {author} {\bibfnamefont {S.}~\bibnamefont {Sharma}},
  \bibinfo {author} {\bibfnamefont {E.}~\bibnamefont {Kandelaki}}, \ and\
  \bibinfo {author} {\bibfnamefont {O.~A.}\ \bibnamefont {Pankratov}},\
  }\href@noop {} {\bibfield  {journal} {\bibinfo  {journal} {Phys. Rev. B}\
  }\textbf {\bibinfo {volume} {81}},\ \bibinfo {pages} {165105} (\bibinfo
  {year} {2010})}\BibitemShut {NoStop}%
\bibitem [{\citenamefont {Abanin}\ \emph {et~al.}(2006)\citenamefont {Abanin},
  \citenamefont {Lee},\ and\ \citenamefont {Levitov}}]{Abanin2006}%
  \BibitemOpen
  \bibfield  {author} {\bibinfo {author} {\bibfnamefont {D.~A.}\ \bibnamefont
  {Abanin}}, \bibinfo {author} {\bibfnamefont {P.~A.}\ \bibnamefont {Lee}}, \
  and\ \bibinfo {author} {\bibfnamefont {L.~S.}\ \bibnamefont {Levitov}},\
  }\href {\doibase 10.1103/PhysRevLett.96.176803} {\bibfield  {journal}
  {\bibinfo  {journal} {Phys. Rev. Lett.}\ }\textbf {\bibinfo {volume} {96}},\
  \bibinfo {pages} {176803} (\bibinfo {year} {2006})}\BibitemShut {NoStop}%
\bibitem [{\citenamefont {Lee}\ and\ \citenamefont
  {Ramakrishnan}(1985)}]{Lee1985}%
  \BibitemOpen
  \bibfield  {author} {\bibinfo {author} {\bibfnamefont {P.~A.}\ \bibnamefont
  {Lee}}\ and\ \bibinfo {author} {\bibfnamefont {T.~V.}\ \bibnamefont
  {Ramakrishnan}},\ }\href {\doibase 10.1103/RevModPhys.57.287} {\bibfield
  {journal} {\bibinfo  {journal} {Rev. Mod. Phys.}\ }\textbf {\bibinfo {volume}
  {57}},\ \bibinfo {pages} {287} (\bibinfo {year} {1985})}\BibitemShut
  {NoStop}%
\bibitem [{\citenamefont {Zhang}\ \emph {et~al.}(2006)\citenamefont {Zhang}
  \emph {et~al.}}]{Zhang2006}%
  \BibitemOpen
  \bibfield  {author} {\bibinfo {author} {\bibfnamefont {Y.}~\bibnamefont
  {Zhang}} \emph {et~al.},\ }\href {\doibase 10.1103/PhysRevLett.96.136806}
  {\bibfield  {journal} {\bibinfo  {journal} {Phys. Rev. Lett.}\ }\textbf
  {\bibinfo {volume} {96}},\ \bibinfo {pages} {136806} (\bibinfo {year}
  {2006})}\BibitemShut {NoStop}%
\bibitem [{\citenamefont {Checkelsky}\ \emph {et~al.}(2009)\citenamefont
  {Checkelsky}, \citenamefont {Li},\ and\ \citenamefont
  {Ong}}]{Checkelsky2009}%
  \BibitemOpen
  \bibfield  {author} {\bibinfo {author} {\bibfnamefont {J.~G.}\ \bibnamefont
  {Checkelsky}}, \bibinfo {author} {\bibfnamefont {L.}~\bibnamefont {Li}}, \
  and\ \bibinfo {author} {\bibfnamefont {N.~P.}\ \bibnamefont {Ong}},\ }\href
  {\doibase 10.1103/PhysRevB.79.115434} {\bibfield  {journal} {\bibinfo
  {journal} {Phys. Rev. B}\ }\textbf {\bibinfo {volume} {79}},\ \bibinfo
  {pages} {115434} (\bibinfo {year} {2009})}\BibitemShut {NoStop}%
\bibitem [{\citenamefont {Giesbers}\ \emph {et~al.}(2009)\citenamefont
  {Giesbers} \emph {et~al.}}]{Giesbers2009}%
  \BibitemOpen
  \bibfield  {author} {\bibinfo {author} {\bibfnamefont {A.~J.~M.}\
  \bibnamefont {Giesbers}} \emph {et~al.},\ }\href {\doibase
  10.1103/PhysRevB.80.201403} {\bibfield  {journal} {\bibinfo  {journal} {Phys.
  Rev. B}\ }\textbf {\bibinfo {volume} {80}},\ \bibinfo {pages} {201403}
  (\bibinfo {year} {2009})}\BibitemShut {NoStop}%
\bibitem [{\citenamefont {Zhao}\ \emph {et~al.}(2010)\citenamefont {Zhao},
  \citenamefont {Cadden-Zimansky}, \citenamefont {Jiang},\ and\ \citenamefont
  {Kim}}]{Zhao2010}%
  \BibitemOpen
  \bibfield  {author} {\bibinfo {author} {\bibfnamefont {Y.}~\bibnamefont
  {Zhao}}, \bibinfo {author} {\bibfnamefont {P.}~\bibnamefont
  {Cadden-Zimansky}}, \bibinfo {author} {\bibfnamefont {Z.}~\bibnamefont
  {Jiang}}, \ and\ \bibinfo {author} {\bibfnamefont {P.}~\bibnamefont {Kim}},\
  }\href {\doibase 10.1103/PhysRevLett.104.066801} {\bibfield  {journal}
  {\bibinfo  {journal} {Phys. Rev. Lett.}\ }\textbf {\bibinfo {volume} {104}},\
  \bibinfo {pages} {066801} (\bibinfo {year} {2010})}\BibitemShut {NoStop}%
\bibitem [{\citenamefont {Kharitonov}(2011)}]{Kharitonov2011}%
  \BibitemOpen
  \bibfield  {author} {\bibinfo {author} {\bibfnamefont {M.}~\bibnamefont
  {Kharitonov}},\ }\href {http://arxiv.org/abs/1103.6285} {\  (\bibinfo {year}
  {2011})},\ \Eprint {http://arxiv.org/abs/1103.6285v1} {arXiv:1103.6285v1}
  \BibitemShut {NoStop}%
\end{thebibliography}
%

\end{document}